\documentclass[11pt,a4paper]{article}
\usepackage{geometry}
\usepackage[utf8]{inputenc}
\usepackage{amsmath,color,titlesec,amsfonts}
\usepackage{graphicx}
\usepackage[lofdepth,lotdepth]{subfig}
\usepackage{tikz}
\usepackage{array}
\usepackage{lscape}
\usepackage[]{graphicx}
\usepackage{multirow}
\usepackage{float}
\restylefloat{table}
\providecommand{\keywords}[1]
{
	\small	
	\textbf{\textit{Keywords---}} #1
}
\allowdisplaybreaks

\title{Analyzes of Algebraic Classification of Higher Dimensional Kundt Geometries with Large $D$ Method }
\author{Pınar KİREZLİ$^{1}$\\ 
$^{1}$Department of Physics, Faculty of Arts and Sciences, Namık Kemal University,Tekirdağ, Turkey \\
pkirezli@nku.edu.tr}

\date{}
\begin{document}
	\maketitle
	\newpage
	\begin{abstract}
In this paper, classification of higher dimensional Kundt geometry is revisited as the dimension of the spacetime $D\rightarrow\infty$. In addition to previous studies, in order to Kundt geometry becomes algebraically special spacetime obligatory conditions are determined. Additionally, Type II, Type III, Type N, Type O and Type D Kundt geometries are explicitly analyzed. Classification of several metrics such as pp-waves, non-gyratonic Kundt metric and VSI spacetime, which are well-known subclasses of Kundt geometry are studied.
	\end{abstract}

 
\section{Introduction}
\label{section1} 
One of the most important classes  of exact spacetimes in general relativity is Kundt spacetime which is introduced by Wolfgang Kundt \cite{Kundt61,Kundt62}. It admits a null geodesic congruence which is non-expanding, twist-free and shear-free and it contains many different vacuum and pure radiation solutions in $D=4$ dimensions such as, pp-waves, plane waves, Bertotti-Robinson, (anti)-Nariai, Plesa$\acute{\text{n}}$ki-Hacyan. Also, Kundt spacetime includes various Petrov types. Kundt waves which are Type N spacetimes representing plane-fronted gravitational waves are commonly investigated \cite{Podolsky:2004qu,Griffiths:2003bk,Ortaggio:2018zze,Podolsky:2007zz}. Further, geodesic motion of Kundt Type III spacetime is analyzed and it is shown that chaotic motion appers under certain conditions \cite{Sakalli:2007ei}. Additionally, Kundt solutions were widely studied with alternative theories of general relativity to understand structure of this spacetime, such as Brans-Dicke theory \cite{Siddhant:2020gkn}, Eddington-inspired Born-Infeld gravity \cite{Chakraborty:2021qxp}, Modified Gravity \cite{Baykal:2015fix}, Einstein-Gauss-Bonnet theory \cite{Svarc:2020fia}. While, explicit extension to higher dimension can be obtained for Kundt geometry \cite{Podolsky:2008ec,Coley:2009ut}, such spacetimes which are sub-classes of Kundt spacetime in $D=4$ dimension, extensions of higher dimensions should be investigated and analyzed.

Higher dimensional solutions in general relativity gained a new perspective with large $D$ expansion method that pioneers are Emparan and et al \cite{Emparan:2013moa,Emparan:2013xia,Emparan:2014cia,Emparan:2015gva,Emparan:2015rva,Andrade:2019edf}. Recently, a review  \cite{Emparan:2020inr} is prepared about general aspects of black holes and effective membrane theories in the large $D$ limit, and some physical problems are discussed in this limit. According to this limitation method, one expects the simplification of Einstein field equations and novel reformulation of the dynamics. In this paper, we will show how the equations are become more simpler for the classification of the Kundt spacetime which was studied by \cite{Kirezli:2022pcu} and its subclasses which are novel classification of the Kundt geometry, as the spacetime dimension $D\rightarrow\infty$.

Petrov classification \cite{Petrov} enables a better comprehension of several aspects of General Relativity in $4$-dimensions \cite{Stephani,Griffits} and it was extended to any $D > 4$ dimensions by \cite{Coley:2004jv}. Explicit and complete classification of the algebraic types and its subtypes of the Kundt geometry based on the Weyl aligned null direction (WAND) for any arbitrary dimension $D>4$ is introduced by \cite{Podolsky:2013qwa}. Although, general aspects of the classification of the Kundt spacetime with large $D$ limitation method found by \cite{Kirezli:2022pcu}, the obligatory conditions are not explicitly studied for Type II, Type III, Type N, Type O and Type D.

In Section \ref{s3} we introduce higher dimensional Kundt spacetime and obtain Weyl scalars with large $D$ limitation method. In Section \ref{s4}, although the Kundt spacetime is not algebraically special, the obligatory conditions are given in order to the spacetime becomes algebraically special. In addition, the necessary conditions for subtypes of Type II are analyzed. Type III, Type N and Type O Kundt geometries for primary WAND $\mathbf{k}$ are discussed in Section \ref{s6} and in Section \ref{s7} Type D geometries for secondary WAND $\mathbf{\ell}$ are studied. Finally, subclasses of the Kundt geometry are explicitly analyzed and the classification of these subclasses are determined in Section \ref{s8}.

\section{Higher Dimensional Kundt Geometries}\label{s3}
Non-twisting, shearfree, non-expanding spacetime is named Kundt spacetime and higher dimensional metric of this spacetime can be written in the form;
\begin{eqnarray}\label{metric}
	ds^2=g_{pq}\left(u,x\right)dx^{p}dx^q+2g_{up}\left(u,r,x\right)dudx^{p}-2du dr+g_{uu}\left(u,r,x\right)du^2 
\end{eqnarray}
where the indeces $p,q..$ count to $2$- $(D-1)$ and $x=x^2,x^3,...x^{(D-1)}$ shorthand of the ($D-2$) spatial coordinates on the traverse space. Other coordinate $r$ is an affine parameter along the optically privileged null congruence ($k=\partial_r$) and the $u=$const. is a null hypersurface. Although the spatial metric coefficient $g_{pq}$ is independent of the parameter $r$, other metric components $g_{up}, g_{uu}$ are functions of $(r,u,x)$. Some of the relations between covariant and contravariant metric components can be written as;
\begin{eqnarray*}
	g^{ur}=-1,~~g^{rp}=g^{pq}g_{uq},~~g^{rr}=-g_{uu}+g^{pq}g_{up}g_{uq},~~g_{up}=g^{rq}g_{pq}.
\end{eqnarray*}

To find out Weyl scalar components of the metric (\ref{metric}), which are necassary to determine algebraic classification of a spacetime, we introduce most natural null frames $\mathbf{k}, \mathbf{\ell}, \mathbf{m_i}$, whose vectors satisfy the normalization conditions $\mathbf{k}.\mathbf{\ell}=-1$, $\mathbf{m_i}.\mathbf{m_j}=\delta_{ij}$ and $\mathbf{k}.\mathbf{k}=\mathbf{\ell}.\mathbf{\ell}=0,~\mathbf{k}.\mathbf{m_i}=\mathbf{\ell}.\mathbf{m_i}=0$. Boosts are defined, according to rescaling of these null frames such as; $\mathbf{k}=\lambda\mathbf{k},~\mathbf{\ell}=\lambda^{-1}\mathbf{\ell},~\mathbf{m_i}=\mathbf{m_i}$, and boost weight which is used to determine the classification of the spacetimes in higher dimensions, are became $+1,-1,0,$ respectively. If Weyl scalar which has $+2$ boost weight, vanishes the null direction of $\mathbf{k}$ becomes a primary Weyl-aligned null direction (WAND). In addition to this, if the all Weyl scalars which have $+1$ boost weight, equal zero, the spacetime becomes algebraically special. The secondary WAND can be defined as null direction of $\mathbf{\ell}$ for fixed $\mathbf{k}$ and the spacetime can be classified more explicitly by secondary WAND.

Weyl scalars of the metric (\ref{metric}) is obtained very uncomplicated as the dimension of the spacetime $D\rightarrow \infty$ (Christoffel symbols, Riemann tensor, Ricci tensor, Ricci scalar and Weyl tensor are calculated in the Apendix \ref{appendixa});
\begin{eqnarray}
	\Psi_{0^{ij}}&=&C_{abcd}k^{a}m^{b}_ik^cm_j^d=0,\\
	\Psi_{1T^{i}}&=&C_{abcd}k^{a}\ell^bk^cm_i^d=m_{i}^p\left(-\frac{1}{2}g_{up,rr}\right),~~\\
	\Psi_{1^{ijk}}&=&C_{abcd}k^{a}m_i^bm_j^cm_k^d=0,\\
	\Psi_{2S}&=&C_{abcd}k^{a}\ell^b\ell^ck^d=\frac{1}{2}g_{uu,rr}-\frac{1}{4}g^{pq}g_{up,r}g_{uq,r},\\
	\Psi_{2T^{ij}}&=&C_{abcd}k^{a}m_i^b\ell^cm_j^d\nonumber\\
	&=&m_i^pm_j^q\bigg(\frac{1}{2}g_{up}g_{uq,rr}+\frac{1}{4}g_{up,r}g_{uq,r}+\frac{1}{2}g_{pn}g^{ms}g_{us,r}~^s\Gamma^n_{~mq}+\frac{1}{2}g_{pn}\left(g^{nm}g_{um,r}\right)_{,q}\bigg),\\
	\Psi_{2^{ijkl}}&=&C_{abcd}m_i^{a}m_j^bm_k^cm_l^d=m_i^pm_j^qm_k^nm_l^mC_{pqnm},\\
	\Psi_{2^{ij}}&=&C_{abcd}k^{a}\ell^b m_i^cm_j^d=m_i^p m_j^q \left(g_{u[p,q]r}+g_{u[p}g_{q]u,rr}\right),\\
	\Psi_{3T^{i}}&=&C_{abcd}\ell^{a}k^b\ell^cm_i^d=m_i^p\left(\frac{1}{4}g_{uu}g_{up,rr}-g_{u[u,p]r}+\frac{1}{2}g^{mn}g_{um,r}E_{np}-\frac{1}{2}g_{up}g_{uu,rr}\right),\\
	\Psi_{3^{ijk}}&=&C_{abcd}\ell^{a}m_i^bm_j^cm_k^d\nonumber\\
	&=&m_i^pm_j^qm_k^m\bigg(-2g_{up}g_{u[q,m]r}+g_{up}g_{u[q}g_{m]u,rr}+g_{up}g_{u[q}g_{m]u,r}\nonumber\\
	&&+g^{\ell s}g_{us,r}~^s\Gamma^n_{~\ell p}g_{u[q}g_{m]n}+\frac{1}{2}g_{up,r}g_{u[q}g_{m]u,r}+g_{u[q}g_{m]n}\left(g^{n\ell}g_{u\ell,r}\right)_{,p}-2g_{pn}g^{sk}E_{k[q}~^s\Gamma^n_{~m]s}\nonumber\\
	&&-g_{pn}\left(g^{rn}g_{u[m,r}\right)_{,q]}-E_{p[m}g_{q]u,r}-2g_{pn}\left(g^{ns}E_{s[q}\right)_{,m]}-g_{pn}g^{rs}~^s\Gamma^n_{~s[q}g_{m]u,r}\bigg),\\
	\Psi_{4^{ij}}&=&C_{abcd}\ell^{a}m_i^b\ell^cm_j^d=m_i^pm_j^q\bigg(-g_{uq}g_{u[u,p]r}+\frac{g_{uu}}{2}\left(g^{ms}g_{us,r}~^s\Gamma^n_{~m(p}g_{q)n}+\left(g^{nm}g_{um,r}\right)_{,(p}g_{q)n}\right)\nonumber\\
	&&+g^{mn}g_{um,r}E_{n(p}g_{q)u}-\frac{g^{rm}}{2}g_{um,r}g_{u(p}g_{q)u,r}+\frac{1}{2}g^{pq}g_{up,r}g_{uq,r}g_{up}g_{uq}\nonumber\\
	&&-g_{pn}\left(-\frac{~^s\Gamma^n_{~sq}}{2}\left(g^{rs}g_{uu,r}+2g^{sm}E_{um}\right)+\left(g^{rn}g_{u[u,r}\right)_{,q]}-2\left(g^{nm}E_{m[q}\right)_{,u]}\right)\bigg).~~
\end{eqnarray}
which were written in order by their boost weight $2,1,0,-1,-2$. These scalars and classification of the Kundt geometry for any arbitrary dimension $D>4$ are discussed in \cite{Podolsky:2013qwa}. The non-zero Weyl scalars do not change with large $D$ method, only the equations become easier.

Further, there are some relations between above Weyl scalars such as;
\begin{eqnarray}
	\Psi_{1T^{i}}&=&\Psi_{1^{k^ki}},\\
	\Psi_{2S}&=&\Psi_{2T^{k^k}},\\
	\Psi_{3T^{i}}&=&\Psi_{3^{k^ki}}.
\end{eqnarray}
Additionally, the symmetric and antisymmetric part of the $\Psi_{2T^{ij}}$ becomes;
\begin{eqnarray}
	\Psi_{2T^{(ij)}}&=&m_i^pm_j^q\bigg(\frac{1}{4}g_{up,r}g_{uq,r}+\frac{1}{2}g_{pn}g^{ms}g_{us,r}~^s\Gamma^n_{~mq}+\frac{1}{2}g_{u(p}g_{q)u,rr}+\frac{1}{2}g_{pn}g_{um,r}g^{nm}_{~~~,q}+\frac{1}{2}g_{u(p,q)r}\bigg)\nonumber\\
	\Psi_{2T^{[ij]}}&=&m_i^pm_j^q\left(\frac{1}{2}g_{u[p}g_{q]u,rr}+\frac{1}{2}g_{u[p,q]r}\right)
\end{eqnarray}
which satisfies the relation $\Psi_{2^{ij}}=2\Psi_{2T^{[ij]}}$. The irreducible components of Weyl scalars are given \cite{Podolsky:2014mpa};
\begin{eqnarray}
	\tilde{\Psi}_{1^{ijk}}&\equiv&\Psi_{1^{ijk}}-\frac{1}{D-3}\left(\delta_{ij}\Psi_{1T^{k}}-\delta_{i k}\Psi_{1T^{j}}\right),\\
	\tilde{\Psi}_{2T^{(ij)}}&\equiv&\Psi_{2T^{(ij)}}-\frac{1}{D-2}\delta_{i j}\Psi_{2S},\\
	\tilde{\Psi}_{2^{ijk\ell}}&\equiv&\Psi_{2^{ijk\ell}}-\frac{2}{D-4}\left(\delta_{i k}\tilde{\Psi}_{2T^{(j\ell)}}+\delta_{j \ell}\tilde{\Psi}_{2T^{(ik)}}-\delta_{i \ell}\tilde{\Psi}_{2T^{(jk)}}-\delta_{j k}\tilde{\Psi}_{2T^{(i\ell)}}\right)\nonumber\\
	&-&\frac{2}{(D-2)(D-3)}\left(\delta_{i k}\delta_{j \ell}-\delta_{i \ell}\delta_{j k}\right)\Psi_{2S},\\
	\tilde{\Psi}_{3^{ijk}}&\equiv&\Psi_{3^{ijk}}-\frac{1}{D-3}\left(\delta_{i j}\Psi_{3T^{k}}-\delta_{i k}\Psi_{3T^{j}}\right).
\end{eqnarray}
where $\delta_{i j}=g_{pq}m_i^{~p}m_j^{~q}$. As the dimension of the spacetime $D\rightarrow\infty$, after the first term of the right hand side vanish and they can be written;
\begin{eqnarray}
	\tilde{\Psi}_{1^{ijk}}\equiv\Psi_{1^{ijk}},~~~~~~~~
	\tilde{\Psi}_{2T^{(ij)}}\equiv\Psi_{2T^{(ij)}},~~~~~~~~
	\tilde{\Psi}_{2^{ijk\ell}}\equiv\Psi_{2^{ijk\ell}},~~~~~~~~
	\tilde{\Psi}_{3^{ijk}}\equiv\Psi_{3^{ijk}}.
\end{eqnarray}

Boost weight $+2$ which is corresponding to $\Psi_{0^{ij}}$, vanishes and as the dimension of the spacetime $D\rightarrow\infty$ the Kundt geometry becomes Type I similar to any arbitrary dimensions $D>4$. Even though one of the Weyl scalar ($\Psi_{1^{ijk}}$) which is corresponding to the $+1$ component of boost weight vanishes, the Kundt geometry is not algebraically special spacetime because of non-zero Weyl scalar ($\Psi_{1T^{i}}$) which is corresponding $+1$ component of boost weight, too. Rest of the paper, first we will analyze algebraically special Kundt geometries without solving field equations and then examine several subclasses of Kundt spacetime. In addition, vanishing Weyl scalar and corresponding (sub)types are summarized in the Table \ref{table1} for primary and secondary WANDs $\mathbf{k}$ and $\mathbf{\ell}$ which is a road map of classification in higher dimensional spacetimes.

\begin{table}[ht]
	\centering
	\begin{tabular}{|c|c|}
		\hline
		\textbf{Types} & \textbf{\begin{tabular}[c]{@{}l@{}}Vanishing \\ Weyl Scalar\end{tabular}}  \\ \hline
		I		& $\Psi_{0^{ij}}$                                                                                                       \\ \hline
		I(a)		& $\Psi_{0^{ij}}$, $\Psi_{1T^{i}}$                                                                                                        \\ \hline
		I(b)	&  	$\Psi_{0^{ij}}$, $\Psi_{1^{ijk}}$                                                                                                      \\ \hline
		II		& $\Psi_{0^{ij}}$, $\Psi_{1T^{i}}$, $\Psi_{1^{ijk}}$                                                                                                      \\ \hline
		II(a)		&                    $\Psi_{0^{ij}}$, $\Psi_{1T^{i}}$, $\Psi_{1^{ijk}}$ , $\Psi_{2S}$                                                   \\ \hline
		II(b)		&   $\Psi_{0^{ij}}$, $\Psi_{1T^{i}}$, $\Psi_{1^{ijk}}$ , $\Psi_{2T^{(ij)}}$                                              \\ \hline
		II(c)		& $\Psi_{0^{ij}}$, $\Psi_{1T^{i}}$, $\Psi_{1^{ijk}}$, $\Psi_{2^{ijk\ell}}$                                                                                 \\ \hline
		II(d)		&  $\Psi_{0^{ij}}$, $\Psi_{1T^{i}}$, $\Psi_{1^{ijk}}$  , $\Psi_{2^{ij}}$                                                                                                 \\ \hline
		III		&  $\Psi_{0^{ij}}$, $\Psi_{1T^{i}}$, $\Psi_{1^{ijk}}$ ,$\Psi_{2S}$, $\Psi_{2T^{(ij)}}$, $\Psi_{2^{ijk\ell}}$, $\Psi_{2^{ij}}$                                                                             \\ \hline
		III(a)		&  $\Psi_{0^{ij}}$, $\Psi_{1T^{i}}$, $\Psi_{1^{ijk}}$,$\Psi_{2S}$, $\Psi_{2T^{(ij)}}$, $\Psi_{2^{ijk\ell}}$, $\Psi_{2^{ij}}$  ,                                                                                                     \\
		& $\Psi_{3T^{i}}$ \\ \hline
		III(b)		&    $\Psi_{0^{ij}}$, $\Psi_{1T^{i}}$, $\Psi_{1^{ijk}}$, $\Psi_{2S}$, $\Psi_{2T^{(ij)}}$, $\Psi_{2^{ijk\ell}}$, $\Psi_{2^{ij}}$,                                                                                                 \\
		&$\Psi_{3^{ijk}}$\\ \hline
		N		&     $\Psi_{0^{ij}}$, $\Psi_{1T^{i}}$, $\Psi_{1^{ijk}}$,$\Psi_{2S}$, $\Psi_{2T^{(ij)}}$, $\Psi_{2^{ijk\ell}}$, $\Psi_{2^{ij}},$                                                                          \\
		& $\Psi_{3T^{i}}$, $\Psi_{3^{ijk}}$                       \\ \hline
		O		&       $\Psi_{0^{ij}}$, $\Psi_{1T^{i}}$, $\Psi_{1^{ijk}}$ ,$\Psi_{2S}$, $\Psi_{2T^{(ij)}}$, $\Psi_{2^{ijk\ell}}$, $\Psi_{2^{ij}},$                                                                                               \\
		&  $\Psi_{3T^{i}}$, $\Psi_{3^{ijk}}$ ,$\Psi_{4^{ij}}$\\ \hline
		I$_i$		& $\Psi_{0^{ij}}$, $\Psi_{4^{ij}}$\\ \hline
		II$_i$		& $\Psi_{0^{ij}}$, $\Psi_{1T^{i}}$, $\Psi_{1^{ijk}}$, $\Psi_{4^{ij}}$\\ \hline
		III$_i$		& $\Psi_{0^{ij}}$, $\Psi_{1T^{i}}$, $\Psi_{1^{ijk}}$,$\Psi_{2S}$, $\Psi_{2T^{(ij)}}$, $\Psi_{2^{ijk\ell}}$, $\Psi_{2^{ij}},$, \\ 
		&$\Psi_{4^{ij}}$\\ \hline
		D	& $\Psi_{0^{ij}}$, $\Psi_{1T^{i}}$, $\Psi_{1^{ijk}}$, $\Psi_{3T^{i}}$, $\Psi_{3^{ijk}}$ ,$\Psi_{4^{ij}}$\\ \hline
	\end{tabular}
	\caption{Algebraic classification of the Kundt geometry for the primary and secondary WANDs $\mathbf{k}, \mathbf{\ell}$ \cite{Podolsky:2014mpa}}. \label{table1} 
\end{table}

\section{Algebraically Special Kundt Geometries}\label{s4}

The spacetime is always Type I(b) because the Weyl scalars $\Psi_{0^{ij}}$ and $\Psi_{1^{ijk}}$ vanish . But it will be algebraicly special when the all boost weight $+1$ of Weyl scalars become zero. The obligatory condition for the Kundt spacetime to becomes algebraically special is vanishing the Weyl scalar $\Psi_{1T^{i}}$. This condition allows us to obtain metric coefficient $g_{up}$ as;
\begin{eqnarray}\label{eqngup}
	g_{up}=	f_p(u,x)r+e_p(u,x).
\end{eqnarray}
and the general Kundt metric becomes;
\begin{eqnarray}
	ds^2=g_{pq}\left(u,x\right)dx^{p}dx^q+2\left(f_p(u,x)r+e_p(u,x)\right)dudx^{p}-2du dr+g_{uu}\left(u,r,x\right)du^2
\end{eqnarray}
where the only metric component of $g_{uu}$ is dependent of the parameter $r$. Additionally, $R_{rurp}$ and $R_{rp}$ vanish without any other conditions. The Weyl scalars become;
\begin{eqnarray}
	\Psi_{2S}&=&\frac{1}{2}g_{uu,rr}-\frac{1}{4}f^pf_p,\\
	\Psi_{2T^{(ij)}}&=&m_i^pm_j^q\bigg(\frac{1}{4}f_pf_q+\frac{1}{2}f^{m}g_{pn}~^s\Gamma^n_{~mq}+\frac{1}{2}g_{pn}f_mg^{nm}_{~~~,q}+\frac{1}{2}f_{(p,q)}\bigg),\\
	\Psi_{2^{ijkl}}&=&m_i^pm_j^qm_k^nm_l^m~^sR_{pqnm},\\
	\Psi_{2^{ij}}&=&m_i^p m_j^q f_{[p,q]},\\
	\Psi_{3T^{i}}&=&\frac{m_i^p}{2}\left(f_{p,u}-g_{uu,pr}+f^{n}E_{np}-\left(f_p r+e_p\right)g_{uu,rr}\right),\label{psi3ti}\\
	\Psi_{3^{ijk}}&=&m_i^pm_j^qm_k^m\bigg(-2\left(f_p r+e_p\right)f_{[q,m]}+\left(f_p r+e_p\right)e_{[q}f_{m]}-g_{pn}g^{rs}~^s\Gamma^n_{~s[q}f_{m]}\nonumber\\
	&&+f^{\ell }~^s\Gamma^n_{~\ell p}\left(f_{[q}g_{m]n}r+e_{[q}g_{m]n}\right)+\frac{1}{2}f_pf_{[m}e_{q]}+\left(f_{[q}g_{m]n}r+e_{[q}g_{m]n}\right)f^{n}_{~,p}\nonumber\\
	&&-2g_{pn}g^{sk}E_{k[q}~^s\Gamma^n_{~m]s}-g_{pn}r\left(f^nf_{[m}\right)_{,q]}-g_{pn}\left(e^nf_{[m}\right)_{,q]}-E_{p[m}f_{q]}-2g_{pn}\left(g^{ns}E_{s[q}\right)_{,m]}\bigg),\label{psi3ijk}\\
	\Psi_{4^{ij}}&=&m_i^pm_j^q\bigg(-\frac{f_qr+e_q}{2}\left(g_{uu,pr}-f_{p,u}\right)+\frac{g_{uu}}{2}\left(f^{m}~^s\Gamma^n_{~m(p}g_{q)n}+f^{n}_{~,(p}g_{q)n}\right)\nonumber\\
	&&+f^{n}\left(E_{n(p}f_{q)}r+E_{n(p}e_{q)}\right)-\frac{g^{rm}f_m}{2}\left(f_{(p}f_{q)}r+e_{(p}f_{q)}\right)+\frac{f^pf_p}{2}\left(f_pr+e_p\right)\left(f_qr+e_q\right)\nonumber\\
	&&-g_{pn}\left[-\frac{~^s\Gamma^n_{~sq}}{2}\left(g^{rs}g_{uu,r}+2g^{sm}E_{um}\right)-2\left(g^{nm}E_{m[q}\right)_{,u]}+\frac{1}{2}\left(g^{rn}g_{uu,r}\right)_{,q}-\frac{1}{2}\left(g^{rn}f_q\right)_{,u}\right]\bigg).\label{psi4ij}~~~~~~
\end{eqnarray} 
where $f^{p}=g^{pq}f_q$. 
\begin{itemize}
	\item The Kundt spacetime becomes algebraically special Type II(a) if the Weyl scalar $\Psi_{2S}=0$. When the condition is applied, we get;
	\begin{eqnarray}\label{sp1}
		g_{uu}=\frac{r^2}{4}f^pf_p+b(u,x)r+c(u,x).
	\end{eqnarray}
	This is same as the result of any arbitrary dimension $D>4$ in \cite{Podolsky:2013qwa}.
	\item The Kundt spacetime becomes algebraically special Type II(b) if the Weyl scalar $\Psi_{2T^{(ij)}}=0$ which reads;
	\begin{eqnarray}\label{sp2}
		\frac{1}{2}f_pf_q+f_{(p,q)}=-g_{pn} \left(f^m~^s\Gamma^n_{~mq}+f_mg^{mn}_{~~~,q}\right).
	\end{eqnarray}
	In this case, it is not possible to obtain an exact result for the metric functions.
	\item  The Kundt spacetime becomes algebraically special Type II(c) if the Weyl scalar $\Psi_{2^{ijkl}}=0$. According to this condition, the Riemann tensor of the (D-2) dimensional traverse space becomes ;
	\begin{eqnarray}\label{sp3}
		~^sR_{pqmn}=0.
	\end{eqnarray} 
	
	\item The Kundt spacetime becomes algebraically special Type II(d) if the Weyl scalar $\Psi_{2^{ij}}=0$ which satisfies;
	\begin{eqnarray}\label{sp4}
		f_{p,q}-f_{q,p}=0.
	\end{eqnarray}
	
\end{itemize}
\section{Type III, Type N and Type O Kundt Geometries}\label{s6}
Kundt spacetime becomes Type III or more special when the above conditions (\ref{sp1}-\ref{sp4}) are satisfied at the same time. We have to analyze the Weyl scalars of $\Psi_{3T^{i}}$ and $\Psi_{3^{ijk}}$ for more special spacetimes with respect to primary WAND $\mathbf{k}$. Further, the classification of Kundt spacetime is summarized in the Table \ref{table2}.
\begin{itemize}
	\item The Type III Kundt geometry is subtype Type III(a) when the Weyl scalar $\Psi_{3T^{i}}=0$ which yields;
	\begin{eqnarray}
		&&f_{p,u}-b_p+\frac{f^n}{2}\left(2e_{[n,p]}+g_{np,u}\right)+\frac{1}{2}f^nf_ne_p=\frac{r}{2}\left[-\left(f^nf_n\right)_{,p}+f^nf_nf_p\right].
	\end{eqnarray}
	Despite the right-hand side of the above equation depend on the coordinate $r$, left hand side independent of the $r$ parameter. As a result ;
	\begin{eqnarray}
		\left(f^nf_n\right)_{,p}&=&f^n f_n f_p\label{type3_1}\\
		b_{,p}&=&f_{p,u}+f^ne_{[n,p]}+\frac{f^n}{2}\left(g_{np,u}+f_ne_p\right),\label{type3_2}.
	\end{eqnarray}
	\item  The Type III Kundt geometry is subtype Type III(b) when the Weyl scalar $\Psi_{3^{ijk}}=0$;
	\begin{eqnarray}\label{type3_3}
		&&r\left[f_pe_{[q}f_{m]}-g_{pn}f^s~^s\Gamma^n_{~s[q}f_{m]}+\left(f^{\ell}~^s\Gamma^n_{~\ell p}+f^n_{,p}\right)f_{[q}g_{m]n}-g_{pn}f^n_{~,[q}f_{m]}\right]\nonumber\\
		&&+e_pe_{[q}f_{m]}+\frac{1}{2}f_pf_{[m}e_{q]}-g_{pn}\left(e^s~^s\Gamma^n_{~s[q}f_{m]}+e^n_{~,[q}f_{m]}\right)+\left(f^{\ell}~^s\Gamma^n_{~\ell p}+f^n_{~,p}\right)e_{[q}g_{m]n}\nonumber\\
		&&-2g_{pn}\left[g^{sk}\tilde{E}_{k[q}~^s\Gamma^n_{~m]s}+\tilde{E}_{s[q}g^{ns}_{~~~,m]}+g^{ns}\tilde{E}_{s[q,m]}\right]-\tilde{E}_{p[m}f_{q]}=0
	\end{eqnarray}
	where $\tilde{E}_{pq}=e_{[p,q]}+\frac{1}{2}g_{pq,u}$. The first line of the equation contains only $r$ parameter and the parenthesis of the first line should be zero for the equation vanishes. This result contains only $f_p, e_p$ and $g_{pq}$ coefficients and the relation between them can be written. Further, the sum of the second and third lines of the equation should be zero which will give some relation between the same coefficients.
	\item The Kundt geometry is type N when the equations (\ref{type3_1}) and (\ref{type3_2}) and $\Psi_{3^{ijk}}=0$ are satisfied simultaneously. 
	\item The Kundt spacetime becomes Type O when the all Weyl scalars vanish. We will not investigate $\Psi_{4^{ij}}=0$ condition which is not easy to analyze but it is written in quadratic form of $A(u,x)r^2+B(u,x)r+C(u,x)$ where the $A,B,C$ are contain $f_q,e_q,b,c$ and $g_{pq}$.
\end{itemize}

\begin{table}[ht]
	\centering
	\begin{tabular}{|c|c|}
		\hline
		\textbf{Types} & \textbf{Obligatory Conditions} \\ \hline
		I  &-                                \\ \hline
		I(a)  & $g_{up}=f_pr+e_p$            \\ \hline
		I(b) &  -                            \\ \hline
		II	                                                                          &  $g_{up}=f_pr+e_p$                               \\ \hline
		II(a)		                                                      &    $g_{uu}=\frac{r^2}{4}f^pf_p+br+c$                         \\ \hline
		II(b)	                                                                          &  $	\frac{1}{2}f_pf_q+f_{(p,q)}=-g_{pn} \left(f^m~^s\Gamma^n_{~mq}+f_mg^{mn}_{~~~,q}\right)   $                           \\ \hline
		II(c)	                                                                             &           	$~^sR_{pqmn}=0$                     \\ \hline
		II(d)		                                                                           &      $f_{[p,q]}=0$                          \\ \hline
		III		                                                                            &       $g_{uu}=\frac{r^2}{4}f^pf_p+br+c$  ,                       \\ 
		&  $	\frac{1}{2}f_pf_q+f_{(p,q)}=-g_{pn} \left(f^m~^s\Gamma^n_{~mq}+f_mg^{mn}_{~~~,q}\right)   $, \\
		&  $~^sR_{pqmn}=0$, \\ 
		&   $f_{[p,q]}=0$ \\ \hline
		III(a)                                                                              &   $	\left(f^nf_n\right)_{,p}=f^n f_n f_p,$\\
		&  $b_{,p}=f_{p,u}+f^ne_{[n,p]}+\frac{f^n}{2}\left(g_{np,u}+f_ne_p\right)    $                         \\ \hline
		III(b)		                                                                        &      Equation (\ref{type3_3})                          \\ \hline
		N	                                                                          &    $	\left(f^nf_n\right)_{,p}=f^n f_n f_p$,\\
		& $b_{,p}=f_{p,u}+f^ne_{[n,p]}+\frac{f^n}{2}\left(g_{np,u}+f_ne_p\right),    $\\
		&         Equation (\ref{type3_3})                       \\ \hline
		O	                                                                         &        $ Ar^2+Br+C=0 $                      \\ \hline
	\end{tabular}
	\caption{Algebraic classification of the higher dimensional Kundt geometries for the primary WAND $\mathbf{k}$ as the dimension of the spacetime $D\rightarrow\infty$.}  \label{table2}
\end{table}
\section{Type D Kundt Geometry}\label{s7}

Kundt spacetime becomes Type D when the Weyl scalars of $\Psi_{0^{ij}},\Psi_{1T^{i}},\Psi_{1^{ijk}},\Psi_{3T^{i}},\Psi_{3^{ijk}},\Psi_{4^{ij}}$ are vanishing. First, in order to necessary condition $\Psi_{1T^{i}}=0$, the metric coefficient of $g_{up}$ is obtained in equation (\ref{eqngup}). By using this metric coefficient, other vanishing Weyl scalars are explicitly given in equations (\ref{psi3ti})-(\ref{psi4ij}) which become as the dimension of the spacetime $D\rightarrow\infty$;
\begin{eqnarray}
	&&f_{p,u}+f^{n}E_{np}-\left(f_p r+e_p\right)g_{uu,rr}=g_{uu,pr},\\
	&&-2\left(f_p r+e_p\right)f_{[q,m]}+\left(f_p r+e_p\right)e_{[q}f_{m]}-g_{pn}g^{rs}~^s\Gamma^n_{~s[q}f_{m]}\nonumber\\
	&&+f^{\ell }~^s\Gamma^n_{~\ell p}\left(f_{[q}g_{m]n}r+e_{[q}g_{m]n}\right)+\frac{1}{2}f_pf_{[m}e_{q]}+\left(f_{[q}g_{m]n}r+e_{[q}g_{m]n}\right)f^{n}_{~,p}\nonumber\\
	&&-g_{pn}r\left(f^nf_{[m}\right)_{,q]}-g_{pn}\left(e^nf_{[m}\right)_{,q]}-E_{p[m}f_{q]}=2g_{pn}\left[g^{sk}E_{k[q}~^s\Gamma^n_{~m]s}+\left(g^{ns}E_{s[q}\right)_{,m]}\right],\\
	&&g_{pn}\left[-\frac{~^s\Gamma^n_{~sq}}{2}\left(g^{rs}g_{uu,r}+2g^{sm}E_{um}\right)-2\left(g^{nm}E_{m[q}\right)_{,u]}+\frac{1}{2}\left(g^{rn}g_{uu,r}\right)_{,q}\right]\nonumber\\
	&&=\frac{\left(f_pr+e_p\right)\left(f_qr+e_q\right)}{2}g_{uu,rr}+\frac{1}{2}f^nE_{nq}\left(f_pr+e_p\right)+\frac{g_{uu}}{2}\left(f^{m}~^s\Gamma^n_{~m(p}g_{q)n}+f^{n}_{~,(p}g_{q)n}\right)\nonumber\\
	&&-\frac{g^{rm}f_m}{2}\left(f_{(p}f_{q)}r+e_{(p}f_{q)}\right)+\frac{f^pf_p}{2}\left(f_pr+e_p\right)\left(f_qr+e_q\right).
\end{eqnarray} 

\section{Examples}\label{s8}
We will investigate the classification of the some spacetime which are subclasses of Kundt spacetime.
\begin{enumerate}
	\item $ds^2=g_{pq}dx^pdx^q+2e_pdudx^p-2dudr+cdu^2$

	This metric corresponding pp-waves which are a subclass of Kundt spacetime with all metric functions are independent of the parameter $r$.  They are naturally become Type II(abd). pp-waves become Type III(a) when the Riemann tensor of traverse space is zero ($~^sR_{pqmn}=0$). This means, Kundt spacetime Type III(a) or more special when the traverse space is flat as the dimension of the spacetime $D\rightarrow\infty$.
	
	Additionally, it will be Type N (and also Type III(b)) for primary WAND $\mathbf{k}$, when the metric coefficients satisfy;
	\begin{eqnarray}
		g^{ns}\left(E_{s[q}\right)_{,m]}=-g^{sk}E_{k[q}~^s\Gamma^n_{~m]s}\label{psi3ijkpp}
	\end{eqnarray}
	and will be Type O with;
	\begin{eqnarray}
		g^{sm}E_{um}~^s\Gamma^n_{~sq}=-2\left(g^{nm}E_{m[q}\right)_{,u]}.\label{psi4ijpp}
	\end{eqnarray}
	On the other hand, it will be Type (II)$_{\text{i}}$ for the secondary WAND $\mathbf{\ell}$, when Weyl scalar $\Psi_{4^{ij}}=0$ which yields equation (\ref{psi4ijpp}). If the traverse universe is flat and equation (\ref{psi4ijpp}) is satisfied, the spacetime becomes Type (III)$_\text{i}$ for the secondary WAND $\mathbf{\ell}$. Finally, the spacetime becomes Type D for the WAND $\mathbf{\ell}$, when the equations (\ref{psi3ijkpp}) and (\ref{psi4ijpp}) are simultaneously provided. pp-waves classification for primary and secondary WANDs for any dimension $D>4$ are shown in Table 3 in reference \cite{Podolsky:2013qwa} and for the spacetime dimension $D\rightarrow\infty$ are discussed in Table 3 in the reference \cite{Kirezli:2022pcu}.
	
	\item $ds^2=g_{pq}dx^pdx^q-2dudr+a (u,x) r^2du^2$
	
	As we set the metric coefficient $g_{up}=0$ the metric take the above form. Physically, this metric is called non-gyratonic Kundt geometry because of $g_{up}$ carries physical information representing a beam  of null radiation with internal spin \cite{Bonnor:1970sb,Frolov:2005in,Krtous:2012qa,Podolsky:2014lpa}. According to above metric, the spacetime automatically becomes Type II(b) and Type II(d). It will be Type II(a) and Type II(c) when the conditions $a=0$ and $~^sR_{pqmn}=0$ are satisfied, respectively. If these conditions are simultaneously fulfilled, the spacetime becomes Type III(a). It will be Type III(b) and Type N, if the case
	\begin{eqnarray}
		\left(g^{ns}g_{s[q,u}\right)_{,m]}=-g^{sk}g_{k[q,u}~^s\Gamma^n_{~m]s}
	\end{eqnarray}
	is ensured. With all above cases, the spacetime becomes Type O when $\left(g^{nm}g_{mq,u}\right)_{,u}=0$ which gives $g^{nm}g_{mq,u}=c(x)$ where $c(x)$ is a constant.

	Also, the classification of the secondary WAND $\mathbf{\ell}$ take into account, the spacetime becomes Type I$_{\text{i}}$ and Type II$_\text{i}$ while the $\Psi_{4^{ij}}$ vanishes which yields;
	\begin{eqnarray}
		-r^2\left[g^{sm}a_{,m}~^s\Gamma^n_{~sq}+\left(g^{nm}a_{,m}\right)_{,q}\right]+r\left[\left(g^{rn}a\right)_{,q}-g^{rs}a ~^s\Gamma^n_{~sq}\right]-\frac{1}{2}\left(g^{nm}g_{mq,u}\right)_{,u}=0.
	\end{eqnarray}
	
	Additionally, the spacetime becomes Type III$_\text{i}$ with large $D$ method when the cases $a=0$, $~^sR_{pqmn}=0$ and $g^{nm}g_{mq,u}=c(x)$ are provided at the same time.
	
	Finally we can say that, it will be Type D when
	\begin{eqnarray}
		&&ra_{,p}=0\leftrightarrow a=a(x),\\
		&&\left(g^{ns}g_{s[q,u}\right)_{,m]}=-g^{sk}\left(~^s\Gamma^n_{~s[m}g_{q]k,u}\right),\\
		&&-r^2\left[g^{sm}a_{,m}~^s\Gamma^n_{~sq}+\left(g^{nm}a_{,m}\right)_{,q}\right]+r\left[\left(g^{rn}a\right)_{,q}-g^{rs}a ~^s\Gamma^n_{~sq}\right]-\frac{1}{2}\left(g^{nm}g_{mq,u}\right)_{,u}=0.
	\end{eqnarray}
	\item $ds^2=\delta_{pq}dx^pdx^q+2\left(f_pr+e_p\right)dudx^p-2dudr+\left(ar^2+br+c\right)du^2$
	
	Kundt spacetime can be written as $g_{uu}$ is quadratic in $r$ and the traverse space $g_{pq}=\delta_{pq}$ is flat. Also, it will define as VSI spacetime when their scalar curvature invariants of all orders vanish. Additionally, the coefficients have to satisfy the conditions that;
	\begin{eqnarray}
		a=\frac{1}{4}f^pf_q,~~~~f_{[p,q]=0},~~~~f_{(pq)}+\frac{1}{2}f_pf_q+\delta_{pn}f^m~^s\Gamma^n_{~mq}=0\nonumber
	\end{eqnarray}
	which make spacetime Type III or more special. Due to this conditions, the spacetime can be called VSI spacetime. 
	
	VSI spacetime becomes Type III(a) when the conditions;
	\begin{eqnarray}
		\left(f^nf_n\right)_{,p}&=&f^nf_nf_p \label{typeıııa}\\
		b_{,p}&=&f_{p,u}f^ne_{[n,p]}+\frac{f^nf_ne_p}{2}\label{typeıııa2}
	\end{eqnarray}
	are satisfied as the dimension of the spacetime $D\rightarrow \infty$.
	
	It becomes Type III(b) when the Weyl scalar $\Psi_{3^{ijk}}$ vanishes which yields;
	\begin{eqnarray}
		&&r\left[f_pe_{[q}f_{m]}-\delta_{pn}f^s~^s\Gamma^n_{~s[q}f_{m]}+\left(f^{\ell}~^s\Gamma^n_{~\ell p}+f^n_{,p}\right)f_{[q}g_{m]n}-\delta_{pn}f^n_{~,[q}f_{m]}\right]\nonumber\\
		&&+e_pe_{[q}f_{m]}+\frac{1}{2}f_pf_{[m}e_{q]}-\delta_{pn}\left(e^s~^s\Gamma^n_{~s[q}f_{m]}+e^n_{~,[q}f_{m]}\right)+\left(f^{\ell}~^s\Gamma^n_{~\ell p}+f^n_{~,p}\right)e_{[q}\delta_{m]n}\nonumber\\
		&&-2\delta_{pn}\left[\frac{\delta^{sk}}{4}\left(e_{[k,q]}~^s\Gamma^n_{~ms}-e_{[k,m]}~^s\Gamma^n_{~qs}\right)+\frac{\delta^{ns}}{2}e_{[m,q]s}\right]-\frac{1}{2}\left(e_{[p,m]}f_q-e_{[p,q]}f_m\right)=0.\label{typeıııb}
	\end{eqnarray}
	
	VSI spacetime is Type N when the equations (\ref{typeıııa})-(\ref{typeıııb}) are simultaneously satisfied.
	
\end{enumerate}

\section{Conclusion}\label{s5}
We analyzed explicit classification of shear-free, twist-free and non-expanding geometries as the spacetime dimension $D\rightarrow\infty$. Eventually, the algebraic classification of the Kundt geometry with large $D$ method, is obtained similar to classification of same spacetime for any dimension $D>4$ but with more simpler equations. Christoffel symbols, Riemann tensors, Ricci tensors and scalar, and Weyl tensor of the Kundt spacetime which are used to calculate Weyl scalars,  were examined in Appendix A as the dimension of the spacetime $D\rightarrow\infty$. Also, coherent with previous studies, this spacetime has been shown to be Type I(b) or more special.

On the other hand, according to our calculations, Kundt spacetime can be algebraically special with respect to primary WAND $\mathbf{k}$, if the metric coefficient $g_{up}$ is a linear function of $r$ which is showed previous works for both any dimensions $D>4$ \cite{Podolsky:2013qwa} and as the dimension $D\rightarrow\infty$ \cite{Kirezli:2022pcu}. In addition, Type III, Type N, Type O and Type D classification of the Kundt geometry with large $D$ method were explicitly analyzed and obligatory conditions were specified, for the first time. 

In addition, subclasses of the well-known Kundt spacetime were investigated by analyzing several metrics without solving field equations. We showed that, the spacetime is Type II(abd) or more special in Example 1 which corresponding pp-waves. If the traverse space flat, it becomes Type III(a) or more special. In Example 2 it is shown that, Kundt geometry without gyratonic matter terms is Type II(ac) or more special.  VSI spacetime which was studied in Example 3 is Type III or more special.  After detection of gravitational waves were announced \cite{LIGOScientific:2016aoc,LIGOScientific:2016sjg}, attention of the researchers to them are incredible increase. Our results may help to understand the structure of them as the dimension of the spacetime $D\rightarrow\infty$. Recently, pp-waves of Type III in $D=4$ dimensional spacetime field equations were solved \cite{Kolar:2021uiu}. They will be generalized to higher dimensional Type III spacetime and our results will be usefull to analyze them as the dimension of the spacetime $D\rightarrow\infty$.

Although classification of the Kundt spacetime with large $D$ method was studied our previous work \cite{Kirezli:2022pcu}, explicit classification of the spacetime were not discussed. Our motivation in preparing this paper was to analyze the classification of Kundt spacetime for both primary and secondary WANDs and to examine classification of subclasses of Kundt geometry in detail, with the large $D$ method. 

In this paper, we obtained the classificaiton of the Kundt geometry without solving Einstien field equations. In the future studies, with large $D$ limit, one can discuss geodesic motion of the Kundt spacetime, or can analyze field equations.

\appendix
\renewcommand{\theequation}{A-\arabic{equation}}
\setcounter{equation}{0}

\section{}\label{appendixa}
The non-zero Christoffel symbols of the general Kundt metric are;
\begin{eqnarray}
	\Gamma^{u}_{~uu}&=&\frac{1}{2}g_{uu,r},\\
	\Gamma^{u}_{~up}&=&\frac{1}{2}g_{up,r},\\
	\Gamma^{r}_{~ur}&=&\frac{1}{2}\left(g^{rp}g_{up,r}-g_{uu,r}\right),\\
	\Gamma^{r}_{~up}&=&\frac{1}{2}\left(-g^{rr}g_{up,r}-g_{uu,p}+2g^{rn}E_{np}\right),\\
	\Gamma^{r}_{~uu}&=&\frac{1}{2}\left(-g^{rr}g_{uu,r}-g_{uu,u}+2g^{rn}E_{un}\right),\\
	\Gamma^{r}_{~pq}&=&+\frac{1}{2}g_{pq,u}-g_{u(p,q)}+g_{un}~ ^s\Gamma^{n}_{~pq},\\
	\Gamma^{r}_{~rp}&=&\frac{-g_{up,r}}{2},\\
	\Gamma^{p}_{~uu}&=&\frac{1}{2}\left(-g^{rp}g_{uu,r}+2g^{pn}E_{un}\right),\\
	\Gamma^{p}_{~ur}&=&\frac{1}{2}g^{pq}g_{uq,r},\\
	\Gamma^{p}_{~uq}&=&\frac{1}{2}\left(-g^{rp}g_{uq,r}+2g_{pn}E_{nq}\right),\\
	\Gamma^{m}_{~pq}&=&~^s\Gamma^{m}_{~pq}~,
\end{eqnarray}
where $^s\Gamma^{n}_{~pq}$ is the Chrisstoffel symbol of the spatial metric $g_{pq}$ and,
\begin{eqnarray}
	E_{pq}&=&g_{u[p,q]}+\frac{1}{2} g_{pq,u},\\
	E_{up}&=&g_{u[p,u]}+\frac{1}{2} g_{up,u}.
\end{eqnarray}
The Rieman tensors of the general Kundt metric are ;
\begin{eqnarray}
	R_{prrq}&=&0~,\\
	R_{ruur}&=& \frac{1}{2}g_{uu,rr}-\frac{1}{4}g^{pq}g_{up,r}g_{uq,r}~,\\
	R_{ruup}&=&g_{u[u,p]r}-\frac{1}{2}g^{mn}g_{um,r}E_{np}+\frac{1}{4}g^{rm}g_{um,r}g_{up,r}~,\\
	R_{rurp}&=&-\frac{1}{2}g_{up,rr}~,\\
	R_{rupq}&=&g_{u[p,q]r}~,\\
	R_{prmq}&=&0~,\\
	R_{pruq}&=&-\frac{1}{2}g_{pn}g^{ms}g_{us,r} ~^s\Gamma^{n}_{~mq}-\frac{1}{4}g_{up,r}g_{uq,r}-\frac{1}{2}g_{pn}\left(g^{nm}g_{um,r}\right)_{,q}~,\\
	R_{pumq}&=&g_{up}g_{u[q,m]r}+E_{p[m}g_{q]u,r}+g_{pn} g^{rs}~^s\Gamma^{n}_{~s[q}g_{m]u,r}+2g_{pn}g^{sk}E_{k[q}~^s\Gamma^{n}_{~m] s}\nonumber\\
	&&+g_{pn}\left(g^{rn}g_{u[m,r}\right)_{,q]}+2g_{pn}\left(g^{ns}E_{s[q }\right)_{,m]}~,\\
	R_{puqu}&=&g_{up}g_{u[u,q]r}-E_{p[u}g_{q]u,r}+\frac{1}{2}g^{rs}g_{uq,r}E_{ps}-g^{s\ell}E_{ps}E_{\ell q}-g_{pn}\left(g^{rn}g_{u[u,r}\right)_{,q]}\nonumber\\
	&&-\frac{1}{2}g_{pn}~^s\Gamma^n_{~sq}\left(g^{rs}g_{uu,r}+2g^{sm}E_{um}\right)+\frac{1}{4}g_{up,r}\left(g_{uu,q}+g^{rr}g_{uq,r}-2g^{rs}E_{sq}\right)\nonumber\\
	&&-2g_{pn}\left(g^{nm}E_{m[q}\right)_{,u]},\\
	R_{pqmn}&=&~^sR_{pqmn}.~~
\end{eqnarray}

Components of the Ricci tensor become;
\begin{eqnarray}
	R_{rr}&=&0,\\
	R_{rp}&=& -\frac{1}{2}g_{up,rr}~,\\
	R_{ru}&=&-\frac{1}{2}g_{uu,rr}+\frac{1}{2}g^{rp}g_{up,rr}+\frac{1}{2}g^{ms}g_{us,r}~^s\Gamma^{q}_{~mq}+\frac{1}{2}\left(g^{mq}g_{um,r}\right)_{,q}~,\\
	R_{uu}&=&-\frac{1}{2}g^{rr}g_{uu,rr}-\frac{1}{4}g^{rm}g^{rp}g_{um,r}g_{up,r}-\left(g^{rp}g_{u[u,r}\right)_{,q]}+\frac{1}{4}g^{pq}g_{up,r}g_{uu,q}-2\left(g^{mq}E_{m[q}\right)_{,u]}\nonumber\\
	&&+\frac{1}{2}g^{rp}g^{mn}g_{um,r}E_{np}-g^{pq}E_{p[u}g_{q]u,r}-g^{pq}g^{s\ell}E_{ps}E_{\ell q}-\frac{1}{2}~^s\Gamma^q_{~sq}\left(g^{rs}+2g^{sm}E_{um}\right)),\\
	R_{up}&=&-\frac{1}{2}g^{rr}g_{up,rr}-g_{u[u,p]r}+\left(g^{rm}g_{u[m,r}\right)_{,p]}-\frac{1}{2}g^{rm}g_{um,r}g_{up,r}+\frac{1}{2}g^{mq}E_{qm}g_{up,r}\nonumber\\
	&&+g^{rs}~^s\Gamma^{m}_{~s[p}g_{m]u,r}-\frac{1}{2}g^{rq}g_{pn}g^{ms}g_{us,r}~^s\Gamma^n_{~mq}+2g^{sk}E_{k[p}~^s\Gamma^m_{~m]s}+2\left(g^{ms}E_s[p\right)_{,m]}\nonumber\\
	&&-\frac{1}{2}g^{rq}g_{pn}\left(g^{nm}g_{um,r}\right)_{,q},	\\
	R_{pq}&=&~^sR_{pq}-\frac{1}{2}g_{up,r}g_{uq,r}-\left(g^{nm}g_{um,r}\right)_{(,q}g_{p)n}+g^{r\ell}~^s\Gamma^n_{~\ell[q}g_{n]p,r}-g^{ms}g_{us,r}~^s\Gamma^n_{~m(p}g_{q)n}\nonumber\\
	&&+g_{pq}g^{rn}g_{un},
\end{eqnarray}

Ricci scalar becomes;

\begin{eqnarray}
	R&=&~^{s}R+g_{uu,rr}-2g^{rp}g_{up,rr}-g^{ms}g_{us,r}~^s\Gamma^{q}_{~mq}-\left(g^{mq}g_{um,r}\right)_{,q}+(D-2)g^{pq}g_{up}g_{un}\nonumber\\
	&&-\frac{g^{pq}}{2}g_{up,r}g_{uq,r}.~~~~~~~~
\end{eqnarray}

We calculate the Weyl tensor of the general Kundt metric as the dimension of the spacetime $D\rightarrow\infty$ which simplifies the results as;
\begin{eqnarray}
	C_{rprq}&=&0,\\
	C_{rpru}&=&-\frac{1}{2}g_{up,rr}~,\\
	C_{prmq}&=&0,\\
	C_{ruru}&=&-\frac{1}{2}g_{uu,rr}+\frac{1}{4}g^{pq}g_{up,r}g_{uq,r},\\
	C_{rpuq}&=&\frac{1}{2}g_{pn}g^{ms}g_{us,r}~^{s}\Gamma^{n}_{~mq}+\frac{1}{4}g_{up,r}g_{uq,r}+\frac{1}{2}g_{pn}\left(g^{nm}g_{um,r}\right)_{,q},\\
	C_{rupq}&=&g_{u[p,q]r},\\
	C_{pqmn}&=&~^sR_{pqmn},\\
	C_{ruup}&=&g_{u[u,p]r}-\frac{1}{2}g^{mn}g_{um,r}E_{np}+\frac{1}{4}g^{rm}g_{um,r}g_{up,r},\\
	C_{upmq}&=&-g_{up}g_{u[q,m]r}-g_{pn}\left(g^{rn}g_{u[m,r}\right)_{,q]}-E_{p[m}g_{q]u,r}-2g_{pn}\left(g^{ns}E_{s[q}\right)_{,m]}\nonumber\\
	&&-g_{pn}g^{rs}~^s\Gamma^{n}_{~s[q}g_{m]u,r}-2g_{pn}g^{sk}E_{k[q}~^s\Gamma^n_{~m]s},\\
	C_{upuq}&=&g_{up}g_{u[u,q]r}-g_{pn}\left(\left(g^{rn}g_{u[u,r}\right)_{,q]}-2\left(g^{nm}E_{m[q}\right)_{,u]}-\frac{~^s\Gamma^n_{~sq}}{2}\left(g^{rs}g_{uu,r}+2g^{sm}E_{um}\right)\right)\nonumber\\
	&&+\frac{1}{4}g^{rr}g_{up,r}g_{uq,r}.~~
\end{eqnarray}


\end{document}